# Single-Crystalline Al/Ge Heterostructure with an Atomically Sharp Commensurate Interface

Jian-Huan Wang[1,2,*], Ding-Ming Huang[1,2], Han Gao[1], Yuan Yao[2], H. Q. Xu[1,3,*] and Jian-Jun Zhang[2,*]

[1]*Beijing Academy of Quantum Information Sciences, Beijing 100193, China*
[2]*Beijing National Laboratory for Condensed Matter Physics and Institute of Physics, Chinese Academy of Sciences, Beijing 100190, China*
[3]*Beijing Key Laboratory of Quantum Devices, Peking University, Beijing 100871, China*

## Abstract

A key challenge in developing Al/Ge heterostructures for quantum applications is Al–Ge interdiffusion. This process is facilitated by grain boundaries in polycrystalline films, which degrades interface quality and impair device performance and reliability. Here, we present epitaxial growth of single-crystalline Al(111) on Ge(111) by molecular beam epitaxy, achieving an atomically flat and sharp interface. At the interface, a commensurate 7-Al-lattice/5-Ge-lattice epitaxial relationship is observed, which dramatically reduces the intrinsic lattice mismatch from 28.4% to ~0.1%. Interestingly, this well-ordered interface does not form below a critical thickness of 3 Å. Instead, Al initially nucleates as random clusters, which then transform into two-dimensional (2D) islands and, as Al deposition further increases, eventually develop into a continuous film. By optimizing the growth parameters, we have achieved an ultra-flat Al film with a surface root-mean-square (RMS) roughness of ~ 0.16 nm and an ultra-thin continuous film with thickness of only 2 nm. These epitaxially grown Al-Ge heterostructures, with their atomically flat surfaces and sharp interfaces, provide a promising platform for studying topological quantum states.

## Introduction

Germanium (Ge), owing to its high hole mobility[1-4], strong spin-orbit coupling (SOC)[5,6], and weak sensitivity to nuclear noise[7], has emerged as a mature platform for developing spin qubit[8,9]. Combining Ge with superconductors—particularly aluminum (Al) that has been extensively utilized and well-studied in quantum information science—holds a great promise for establishing a versatile material platform for diverse quantum computing research[10]. Recent studies have demonstrated significant progress in Al/Ge hybrid structures fabricated via *ex situ* deposition and annealing techniques[11-14]. Josephson junctions with high transparency have been realized across several Ge platforms, including Si-Ge core/shell nanowires, Ge/SiGe heterostructures, and planar Ge hut nanowires.[15-17] Furthermore, gate-tunable superconducting qubits made in Ge systems also have been reported.[18-20]

In superconductor-semiconductor hybrid systems, the crystallinity of the superconductor film (single-crystalline or polycrystalline) has not been a primary concern. However, in the Al-Ge system, strong interdiffusion between Al and Ge poses a critical challenge. Grain boundaries in polycrystalline Al films act as low-energy diffusion pathways, thereby significantly reducing the onset temperature of interdiffusion to as low as 180 °C[12,21]. This interdiffusion severely impacts those Josephson junction devices with channel lengths of only tens to hundreds of nanometers, and also leads to a poor reproducibility. It is because the interdiffusion causes variations in the Ge channel length, potential p-type doping of the Ge channel, and the formation of a gate-insensitive AlGe alloy segment in the channel[21]. Furthermore, Al/Ge hybrids are predicted to be a promising platform for hosting topologically protected Majorana zero modes[22,23]. The prospect of exploring such exotic states, in turn, imposes exceptionally stringent requirements on the quality of the Al/Ge interface.

Molecular beam epitaxy (MBE) is widely recognized as a critical approach for obtaining heterostructures with atomically sharp interfaces. MBE-grown Al/III-V hybrid nanowires and heterostructure[24-26] exhibit sharp interfaces and hard superconducting gaps, providing an ideal platform for exploring Majorana zero modes. Furthermore, MBE has also been successfully employed in the study of single-crystal Al/Si heterostructures[27-30]. However, to the best of our knowledge, MBE growth in the Al-Ge system has rarely been reported so far. In this work, we report a method to epitaxially grow Al/Ge heterostructures using a two-step, *in situ* MBE process. This method produces single-crystalline Al(111) films on Ge(111), entirely free of any grain boundaries. Between the face-centered cubic (FCC) -structured Al and the diamond-structured Ge, near single atomic-layer sharp interfaces are realized. At the interfaces, a commensurate epitaxial relationship of a 7-Al-atom-lattice to a 5-Ge-atom-lattice is observed, which reduces their lattice mismatch to as low as 0.1%. Through systematic optimization of growth parameters, atomically flat Al films have been observed by enhancing adatom migration, and continuous Al films with thickness of only 2 nm have been obtained by suppressing adatom migration.

**Experiments**

The Al/Ge heterostructures were grown on commercial epi-ready Ge(111) wafers using a dual-chamber solid-source MBE system with a base pressure of $1\times10^{10}$ mbar. Initially, a 60-nm-thick Ge layer was grown on the substrate in the SiGe MBE chamber. After the Ge growth, the samples were cooled down and *in situ* transferred to the III-V MBE chamber for Al deposition. Two optimized growth recipes were developed. In Recipe A, Al was deposited at a fixed substrate temperature of 40 °C and a rate of 1 Å/s. Recipe B involved Al deposition at a temperature below 0 °C (below the thermocouple detection value) and a growth rate of 0.37 Å/s. It should be noted that the Al growth rates were determined by reflection high-energy electron diffraction (RHEED) oscillations measured during the growth of AlAs. To ensure thickness uniformity of the film, the samples were rotated at 5 r/min during the growth. Consequently, there exists a small systematic error between the actual thickness and the set thickness of the films, but it is less than 5%. (Refer to the Growth Methods section in the Supporting Information for details.)

The film growth is monitored in real time using RHEED. The surface morphology was characterized using atomic force microscope (AFM) operating in tapping mode under ambient conditions. Atomic-scale surface structure and early-stage Al film morphologies were examined by *in situ* scanning tunneling microscopy (STM) at room temperature. The crystal structure of the Al/Ge heterostructure and their interfacial properties were characterized by X-ray diffraction (XRD) and Cs-corrected scanning transmission electron microscopy (STEM), respectively.

**Results**

Figure 1(a) shows a typical AFM image of the as-grown Ge layer, revealing an atomically smooth surface with visible atomic steps. These atomic steps meander along the [110] crystallographic direction with an average spacing of ~200 nm. The atomic-scale structure of this surface is resolved by STM displayed in the inset of Fig. 1(a), which shows a dominant c(2×8) reconstruction. Subsequently, the deposition of a 20-nm-thick Al film using Recipe A results in continuous growth of two-dimensional films, yielding an atomically flat surface as shown in Fig. 1(b). This flat film exhibits a peak-to-valley height of ~1.6 nm and a root-mean-square (RMS) roughness as small as ~0.164 nm over a $5\times5\ \mu m^2$ area — comparable to the RMS roughness of ~0.132 nm measured on the underlying Ge epilayer in Fig. 1(a). The inset STM image in Fig. 1(b) further resolves the sixfold rotational symmetry characteristic of an Al(111) surface.

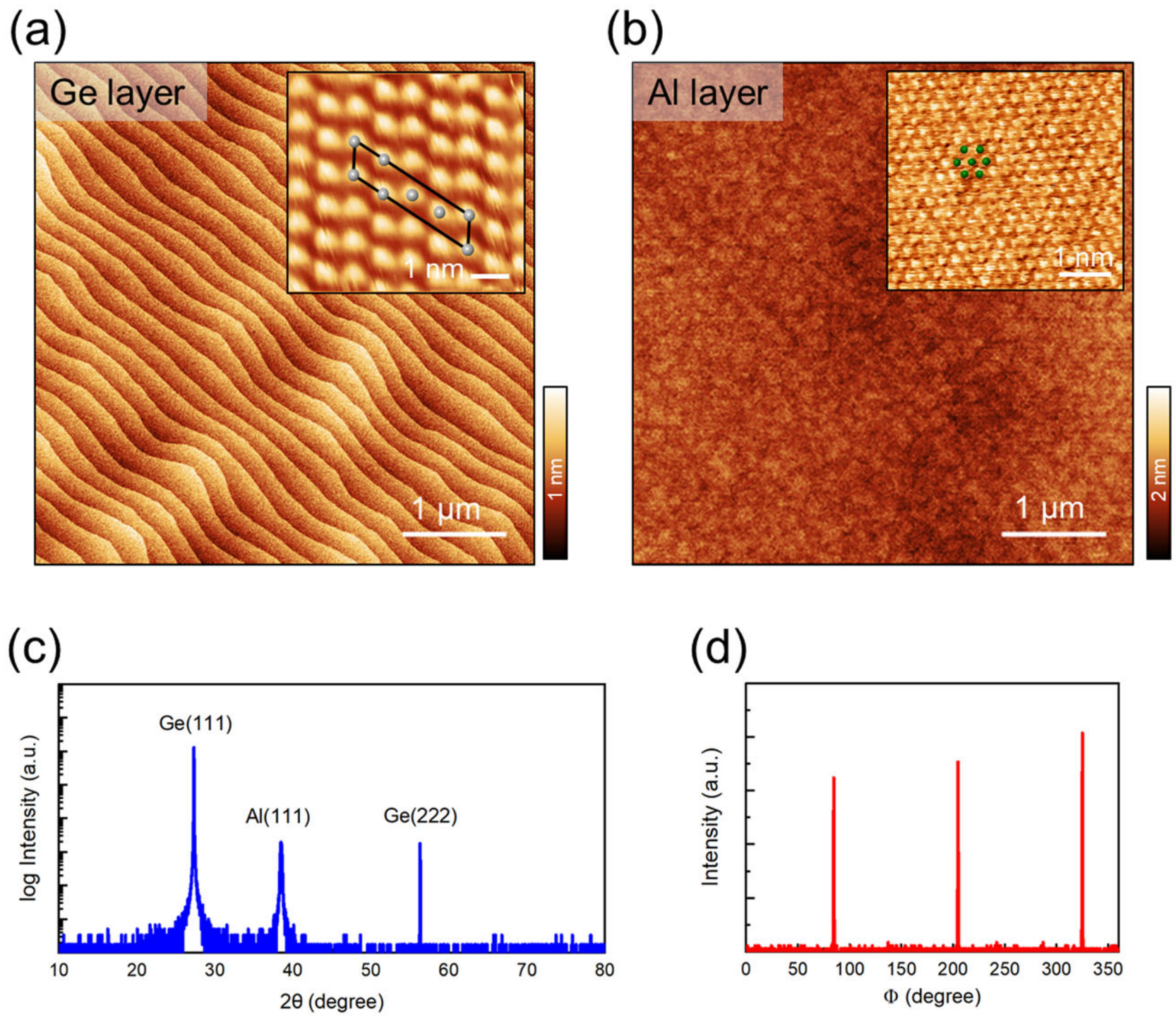

**Figure 1. Morphology and crystallinity of the Al/Ge heterostructure grown by Recipe A.** (a,b) AFM images (5 μm × 5 μm) of the as-grown Ge and the Al epilayer. The insets in (a,b) show the STM images of the corresponding atomic-scale structures. (c,d) XRD spectra taken with symmetric ω-2θ scan and asymmetric Φ-scan of the Al/Ge heterostructure.

A wide XRD ω-2θ scan was performed on the as-grown Al/Ge heterostructure to confirm its crystalline structure, as shown in Fig. 1(c). Three dominant diffraction peaks are observed at 27.28°, 38.47° and 56.29°, which are indexed to the Ge(111), Al(111) and Ge(222) planes, respectively. According to Bragg's law, the measured interplanar spacing of Al(111) was determined to be 2.340 Å, corresponding to a lattice constant of 4.053 Å. This value shows excellent agreement with the theoretical value of bulk Al (4.050 Å), indicating that the 20 nm thick Al epilayer is strain relaxed. Despite the ω-2θ spectrum only displays one peak for Al(111), two equivalent domains with inverted atomic stacking sequences, i.e., Al(111) and Al($\bar{1}\bar{1}\bar{1}$), may coexist, potentially resulting in twinning defects. This twinning tendency has been reported in the previous Al/Si systems[31, 32]. As FCC materials have threefold rotational symmetry along the [111]-axis, the presence of these twinning defects can be identified by asymmetric rotational scans (Φ-scans) with respect to surface normal, appearing as six diffraction peaks. Figure 1(d) shows such asymmetric Φ-scan, with 2θ set towards Al(100) peak and the substrate tilted to 54.7°. Only three diffraction peaks are observed over a full Φ-rotation (360°), confirming the single-crystalline of the film.

Figure 2(a) shows the cross-sectional bright-field TEM image of the Al/Ge heterostructure, distinctly resolving the Ge epilayer, Al film, and native Al oxide. The interface between the 60-nm-thick Ge epilayer and the substrate is invisible, indicating effective removal of surface contaminants through the degassing/deoxygenation process as well as the high-quality Ge epitaxial growth. In contrast, the Al/Ge interface exhibits sharpness with a straight contour that maintain the atomic flatness of the underlying Ge layer, which indicates the suppression of the potential Al-Ge interdiffusion. We observe the nominal 20-nm Al film developed ~3.5-nm native oxides at the surface, leaving residual Al film thicknesses of ~18.5 nm under ambient conditions. We attribute this nearly 10% deviation in total film thickness, on the one hand, to the systematic error of the RHEED rate calibration discussed above, and on the other hand, to the thickness increase resulting from surface oxidation of the Al film. The inset in Fig. 2(a) shows the electron diffraction pattern for the heterostructure along the $[1\bar{1}0]$ zone axis of the substrate. The Al and Ge diffraction spots exhibit coincident angular arrangement with no relative rotation, confirming that the epitaxial relationship is Al(111)$[1\bar{1}0]$ // Ge(111)$[1\bar{1}0]$.

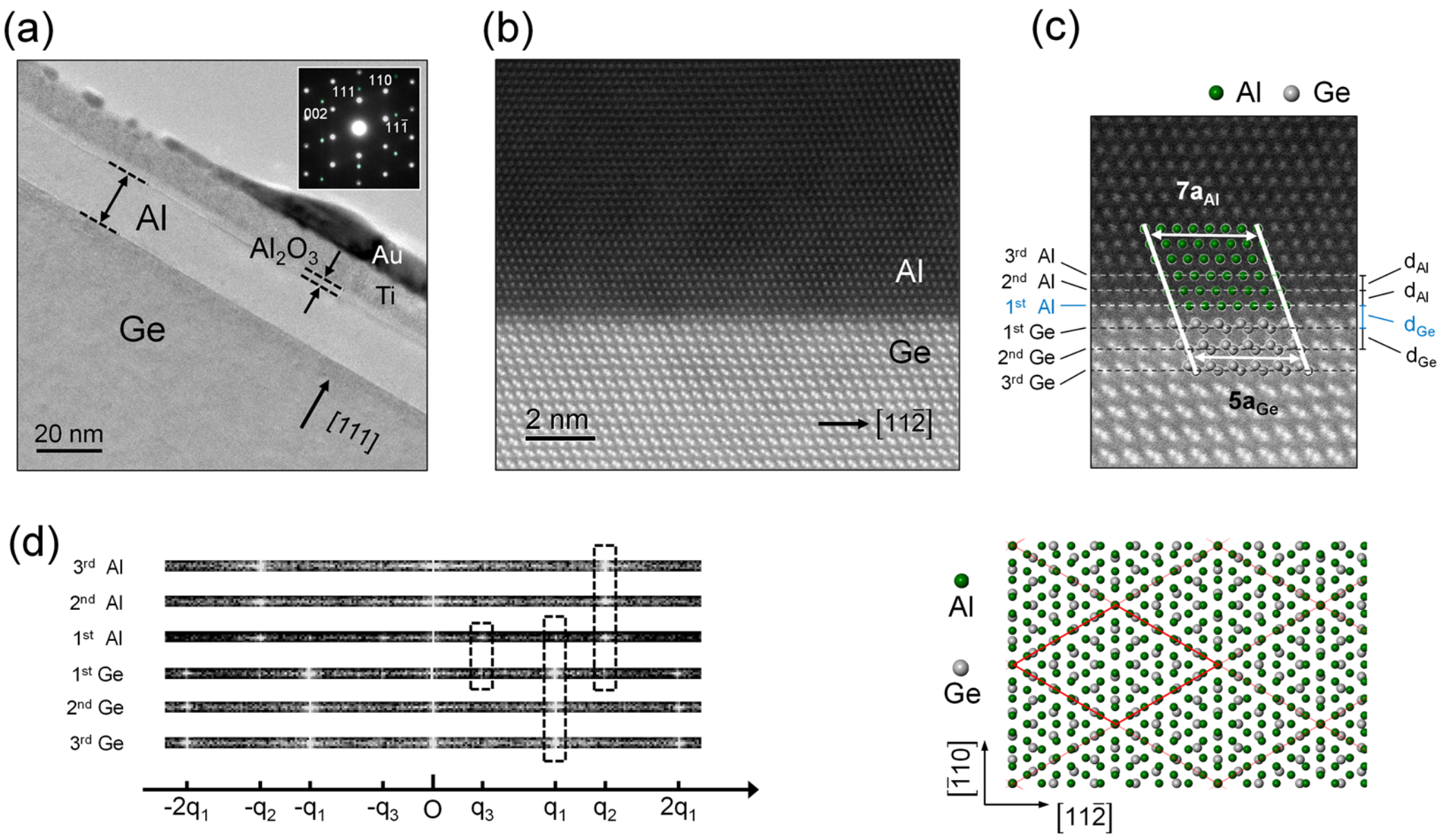

**Figure 2. TEM analyses of the Al/Ge heterostructure.** (a) Low-magnification bright-field TEM image of the Al/Ge heterostructure in a cross-sectional view. The distinct layers of Ge, Al, native Al oxide, and the protective capping (Ti/Au) are labeled. The inset shows an electron diffraction pattern near the Al/Ge interface, where the Ge and Al spots are marked in gray and green, respectively. (b) HAADF-STEM image demonstrating an atomically sharp interface of the Al/Ge heterostructure. (c) Atomic-resolution HAADF-STEM image that clearly reveals the commensurate epitaxial relationship of a 7-Al-atom-lattice to a 5-Ge-atom-lattice at the interface. (d) Schematic of the in-plane atomic arrangement at the Al/Ge interface. A representative supercell, defined by symmetry, is highlighted by the red rhombus. (e) Quasi-1D fast Fourier transform (FFT) patterns from the three interfacial Al and Ge atomic layers, showing the reciprocal lattice points of Ge ($q_1$) and Al ($q_2$), and the difference vector $q_3$ arising from the interaction between Al and Ge atom at the interface.

The high-angle annular dark-field (HAADF) STEM image in Fig. 2(b) resolves the cross-sectional Al/Ge interface, where bright spots correspond to Ge atoms in the diamond cubic structure, while spots with dark contrast represent Al atoms in the FCC structure. An atomically sharp interface between diamond-structured Ge and FCC-structured Al is observed, devoid of any interface compound or amorphous layer. At the interface, we further identify a lateral periodic superlattice consisting of commensurately matched 7-Al to 5-Ge atomic planar lattice, as labeled in Fig. 2(c). For bulk Al and Ge, their corresponding lattice constants are 3.307 Å and 4.620 Å, resulting in a substantial lattice mismatch of 28.4%. The commensurate interface, however, exhibits a drastically reduced mismatch of only ~0.1%. The interlayer distances near the interface were also measured in Fig. 2(c). What we observe is that the spacing between the $1^{st}$ Al and the bottom atom plane of the $1^{st}$ Ge approaches the interplane distance of Ge(111), suggesting the presence of interfacial coupling between Al and Ge atoms.

Figure 2(d) schematically illustrates a rhombic supercell at the Al/Ge interface from a symmetry perspective, with a length of 1.2 nm along the $[1\bar{1}0]$ direction and a length of 2.0 nm along the $[11\bar{2}]$ direction. Through *in situ* STM characterization, we recently observed Moiré patterns on the surface of 1-25 nm thick Al films. The visualizing of interfacial signal in STM surface imaging can be well explained using a Fabry-Perot interferometer model, and the Moiré pattern is attributed to the atomically resolved lateral changes of electron reflectivity at the Al/Ge interfaces, which is modulated by the specific interface lattice.[33] To obtain more direct evidence, we performed a quasi-1D Fast Fourier Transform (FFT) on those three interfacial layers of Al and Ge to extract their periodic characteristics. The corresponding FFT patterns are shown in Figure 2(e). We note that the regions for FFT analysis in the STEM image were carefully selected to avoid artifacts from adjacent atomic layers, as detailed in Fig. S1. The reciprocal lattice points of Ge and Al are denoted as $q_1$ and $q_2$, respectively, and their difference vector is defined as $q_3=q_1-q_2$. We find that the $2^{nd}$ and $3^{rd}$ layers of both Al and Ge exhibit only their intrinsic lattice periodicities. In contrast, the $1^{st}$ layers of both Al and Ge clearly show the presence of $q_3$. A corresponding spot at the $q_3$ position is also resolved in the FFT of our STM measurements, confirming that the Moiré pattern arises from interfacial interactions. We propose that the appearance of this difference vector reflects observable modulation in interfacial coupling. Additionally, no signature of the Ge surface reconstruction (i.e., $q_1/2$) is detected in the FFT patterns.

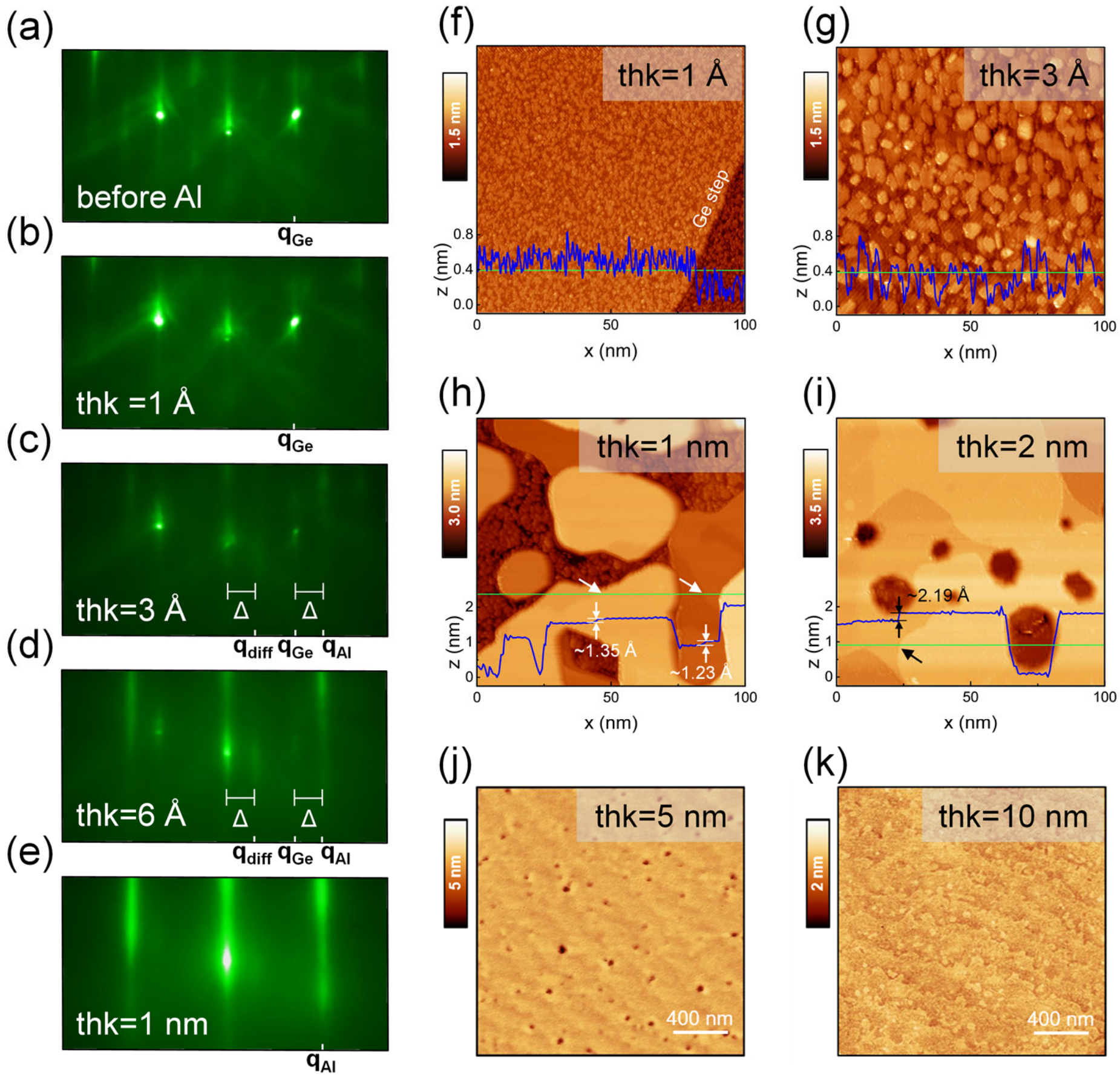

**Figure 3. Al deposition on Ge(111).** (a-e) RHEED patterns acquired before (a) and during Al deposition, at thicknesses of 1 Å (b), 3 Å (c), 6 Å (d), and 1 nm (e). For Al thicknesses between 3 Å (c) and 6 Å (d), the RHEED patterns clearly show the coexistence of Ge×1 and Al×1, accompanied by a set of satellite streaks at a position of $q_{diff} = q_{Al} - q_{Ge} = \Delta$. (f-g) *In situ* STM images (100 nm × 100 nm) of the surface after Al deposition, at thicknesses of 1 Å (f), 3 Å (g), 1 nm (h), and 2 nm (i). The blue lines represent height profiles measured along the paths defined by the green lines. Steps with sub-ML heights and single-ML steps were identified from the profiles in (h) and (i), which are marked by white and black arrows, respectively. (j,k) AFM images (2 μm × 2 μm) after 5 nm and 10 nm Al film growth, respectively.

We next characterize the evolution of Al on Ge(111). Figure 3(a-e) presents a series of RHEED patterns acquired before and during Al deposition, at thicknesses of 1 Å, 3 Å, 6 Å, and 1 nm. Figure 3(f-i) shows the STM images at various Al coverages. In contrast to the well-ordered commensurate Al/Ge interface revealed by STEM images shown in Fig. 2(b,c), the initially deposited Al forms randomly 3D clusters on Ge terraces, as shown in Fig. 3(f). These clusters progressively coalesce and eventually wet the entire surface with Al growth. RHEED measurements show that the original ×2 streaks of Ge vanish immediately upon Al deposition, leaving only an attenuated Ge×1 pattern originating from the underlying Ge layer (See Fig. 3(a,b)). At a thickness of 3 Å, small flat 2D islands form, as shown in Fig. 3(g). Concurrently,

Al×1 pattern becomes faintly visible in the RHEED (Fig. 3(c)). With further Al deposition, the Al×1 intensifies and eventually becomes dominant, whereas the Ge×1 pattern progressively weakens and nearly vanishes at a thickness of 1 nm (Fig. 3(e)). Additionally, there is a set of satellite streaks at the position $q_{diff}=q_{Al}-q_{Ge}$ in the RHEED patterns. These streaks appear simultaneously with the Al×1 and vanish synchronously with the Ge×1.

Figure 3(h) shows the morphology after depositing 1 nm of Al. At this stage, large-area 2D islands with steep edges form randomly, while a cluster-like layer reappears in the regions between these islands. Compared to the morphology shown in Fig. 3(g), this evolution reflects a pronounced process of multilayer atomic mass transfer. The meandering shape of these large islands further reveals that the lateral growth dominate at this stage. In addition, we also see new 2D islands nucleate on terraces. It is noteworthy that the steps with sub-ML heights are observed on the plateau of these islands (the white arrow in Fig. 3(h)), exhibiting a height of approximately 1.2–1.4 Å. These values slightly exceed the height difference of one monolayer between Al and Ge (0.92 Å). A relevant STM study by Y. Jiang et al[28] on Al/Si(111)-$\sqrt{3}\times\sqrt{3}$ revealed that quantum size effects drive the formation of a dilute 1.5×1.5 intercalated layer structure (interlayer spacing of ~3.5 Å) in the initial few MLs of Al films. With increasing Al coverage, strain induces a phase transition from this dilute intercalated structure back to a normal stacking of Al(111)-1×1 layers, where the interlayer spacing is ~2.3 Å. The height difference between the coexisting dilute and normal stacking layers is expected to be ~1.2 Å, which is in agreement with the steps with sub-ML heights in our measurements. Furthermore, the almost absence of such sub-ML-height steps at the thickness of 2 nm is also consistent with the phase transition mechanism proposed by Y. Jiang et al[28].

Upon reaching a thickness of 2 nm, the initially dispersed 2D islands grow into a continuous film, though some pits remain on the surface. Subsequent growth then changes to a layer-by-layer mode, resembling homoepitaxy. We employed *ex situ* AFM to evaluate the overall surface flatness after 5 nm and 10 nm thick Al films, as shown in Fig. 3(j,k). The surface after 5 nm film exhibits some pits with a density of ~$2.5\times10^{9}$ $cm^{-2}$. Nevertheless, the surface maintains a good flatness, and the buried Ge steps can be clearly resolved. The spacing and orientation of these steps closely match those of the pristine Ge substrate surface before Al deposition (Fig. 1(a)), further confirming the formation of a high-quality Al epilayer and an atomically sharp Al/Ge interface. When the thickness increases to 10 nm, these pits are fully disappeared, resulting in a continuous and flat film.

The studies performed at different Al deposition temperatures reveal that the single-crystalline Al films obtained on Ge(111) are, in fact, metastable. Figure 4 presents AFM images of 20-nm-thick Al films deposited at 1 Å/s at different temperatures. We find that single-crystal films form only at low temperatures, i.e. ≤ 60 °C, whereas temperatures above 100 °C yield polycrystalline grains. Within the single-crystal regime (≤60 °C), the smoothest surface is achieved at 40 °C (Recipe A), as previously mentioned. Deviation from this optimal temperature—whether increasing it (up to 60 °C) to enhance adatom mobility or decreasing it (below 40 °C) to restrict migration—still results in single-crystal films but with increased surface roughness, as shown in Fig. 4(c) and Fig. 4(e–f), respectively.

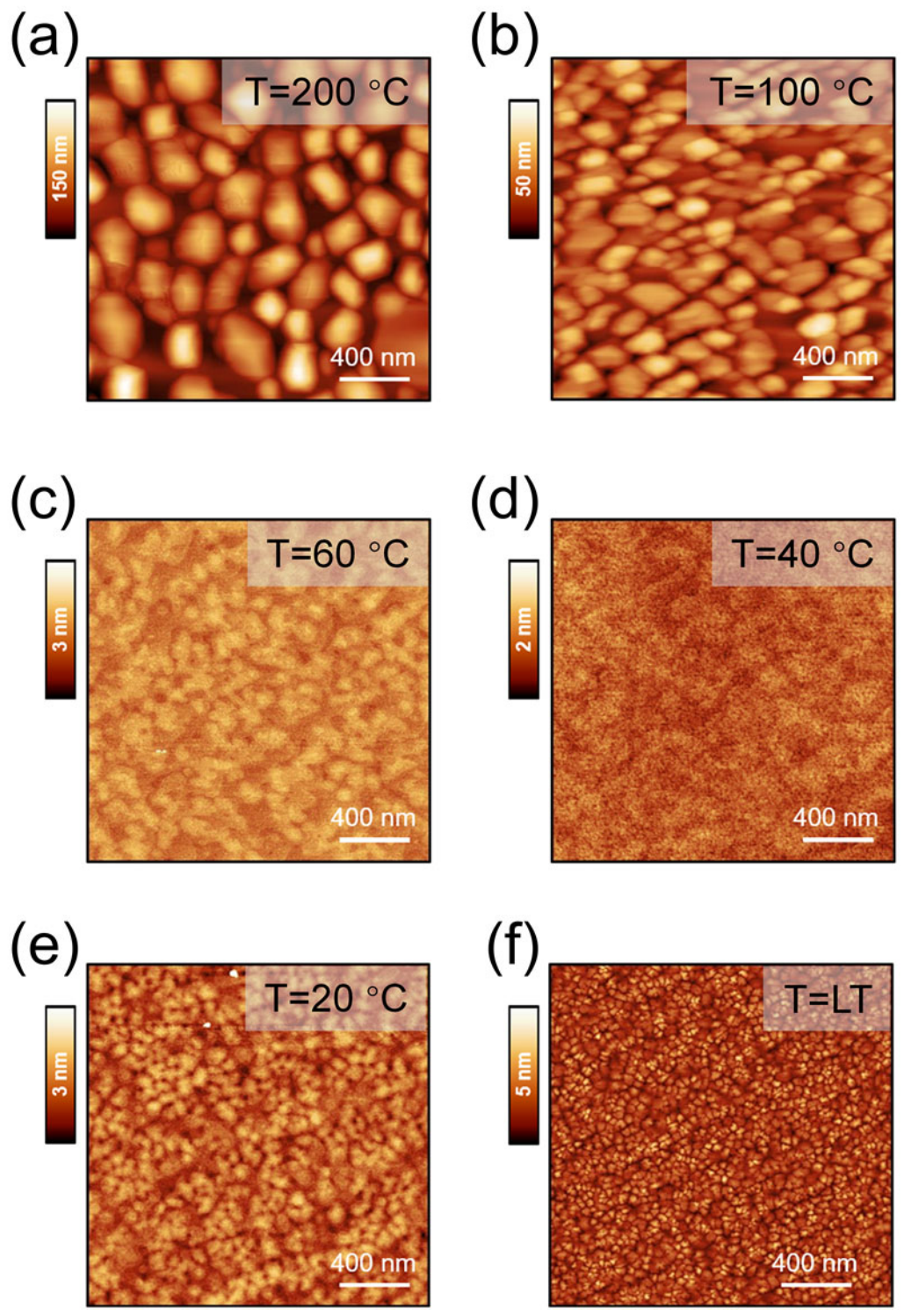

**Figure 4. Effect of growth temperature on the surface morphology of the Al film.** (a-f) AFM images (2 μm × 2 μm) showing 20 nm-thick Al films grown at 200 °C, 100 °C, 60 °C, 40 °C, 20 °C and the low temperature conditions (LT), respectively.

Since low-temperature growth is a common route for realizing high-quality superconductor-semiconductor heterostructures, particularly Al/III-V systems, we next optimize Al film growth under the low-temperature conditions by employing a reduced deposition rate of 0.37 Å/s (Recipe B; see Supporting Information, Fig. S3). RHEED and STM observations indicate that the first Al monolayer still forms 3D clusters, while Al(111) small islands develop in subsequent layers. Owing to reduced adatom migration at low temperature, these small islands then persistently dominate the surface, resulting in high surface flatness even at low Al coverages. AFM images in Fig. 5(a,b) of 2 nm and 5 nm thick Al films confirm that Recipe B yields continuous, pit-free films at a thickness of just 2 nm, which maintain low roughness with further deposition. Notably, the atomic steps of the underlying Ge epilayer remain clearly resolvable through 2 nm, 5 nm, and even 20 nm thick Al films (Fig. 5(a,b) and Fig. S3(c)). These results demonstrate not only atomic flatness at the Al surface but also minimal atomic interdiffusion at the Al/Ge interface in these low-temperature-grown heterostructures.

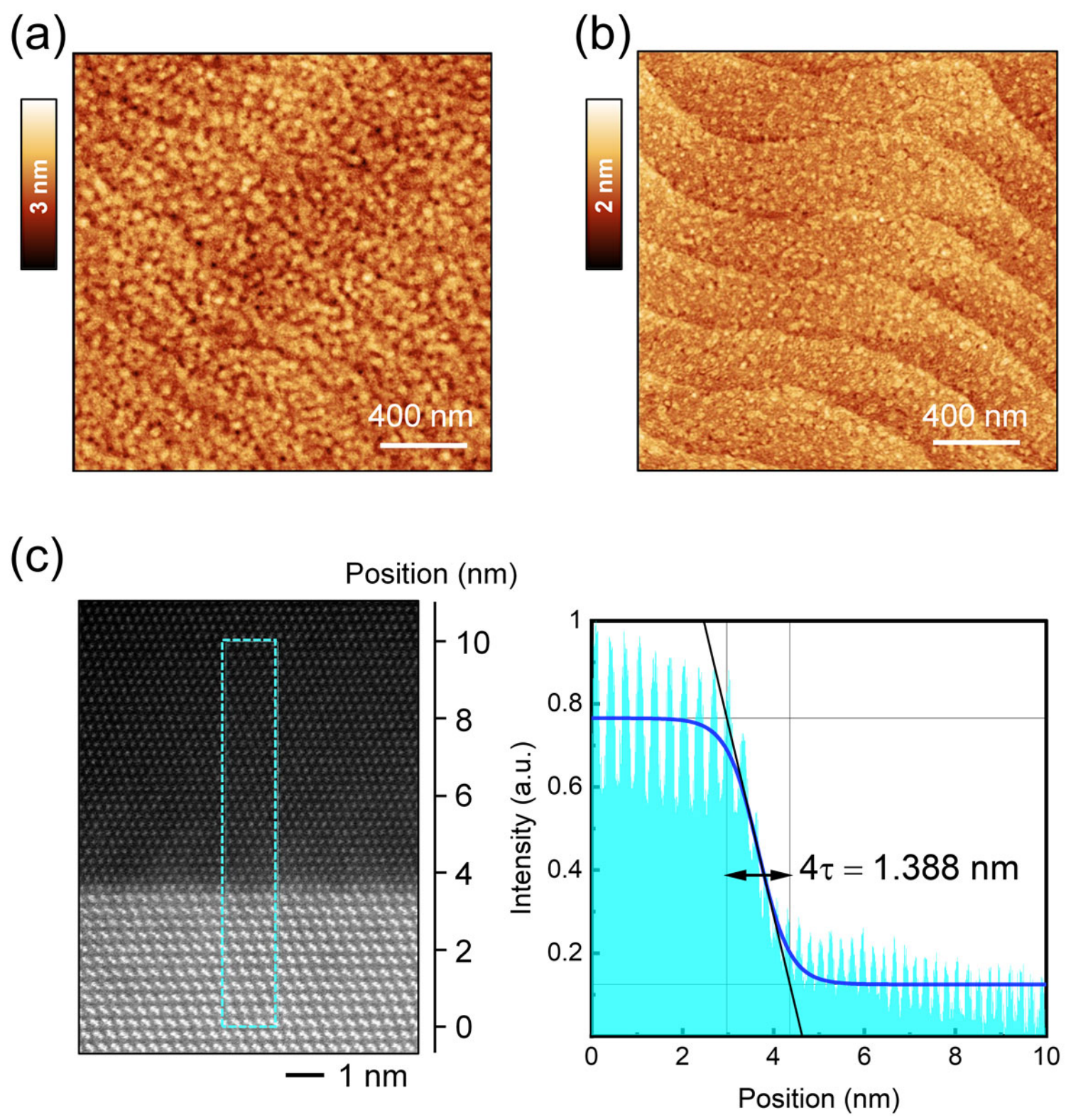

**Figure 5. Characterization of the Al film grown at low temperature (Recipe B).** (a,b) AFM images (2 μm × 2 μm) of the sample surface after 2 nm and 5 nm Al deposition, respectively. (c) Cross-sectional HAADF-STEM image of the Al/Ge interface (left) and its corresponding intensity profile (right) from the cyan-dashed region. The thick blue curve in HADDF intensity profile represents the sigmoid fit, where the fitted parameter 4τ defines the interface width.

The left panel of Fig. 5(c) presents an atomic-resolution HAADF-STEM image of the Al/Ge heterostructure grown via Recipe B, revealing the formation of an epitaxial Al film with an atomically sharp commensurate interface even at a substantially reduced growth temperature. These interfacial characteristics exhibit identical crystallographic alignment to that shown in Fig. 2(b-c). Leveraging the high sensitivity of HAADF intensity to atomic number (Z), the interface width, the value which is often used to quantify the sharpness of a heterogenous material interface, was extracted though fitting the normalized intensity profile with a sigmoid function[4,34] $I(z) = \left(I_{Ge}^0 - I_{Al}^0\right)/\{1 + \exp[(z - z_0)/\tau]\} + I_{Al}^0$. Here, $I_{Ge}^0$ ($I_{Al}^0$) is the average intensity of Ge (Al) layer, $z_0$ is the inflection point of the interface, and the interface width is defined to be where the tangent at position $z_0$ crosses the horizontal asymptotic limits $I = I_{Ge}^0$ and $I = I_{Ge}^0$, corresponding to 4τ (please refer the black guidelines in the right panel of Fig.

5(c)). From the fits shown in the right panel of Fig. 5(c), we determine that the low-temperature grown Al/Ge junction exhibits an interface width of only 1.388 nm. While the Recipe A-grown counterpart under identical fitting methods, shows a slightly wider interface width of 1.534 nm (quantified by 4τ, see Fig. S2).

## Conclusion

In summary, we have successfully realized single-crystalline Al/Ge heterostructures with atomically sharp and commensurate interfaces on Ge(111) wafers via MBE. Prior to Al deposition, *in situ* STM confirmed a well-defined c(2×8) reconstruction on the Ge epilayer surface, whereas the as-grown Al film exhibited an unreconstructed Al(111). Atomic-resolution HAADF-STEM reveal a 7-Al-lattice/5-Ge-lattice commensurate interface, which effectively reduces the in-plane lattice mismatch to merely 0.1%. Quasi-1D FFT analysis identify a distinct difference vector $q_3$ localized at the first atomic layers of Al and Ge, suggesting the presence of a detectable coupling across the interface. Growth temperatures exceeding 100 °C produce only rough, polycrystalline grains. At the growth temperature ≤ 60 °C enables the formation of continuous single-crystal films. Finally, an ultra-flat Al film with a surface RMS roughness of only 0.164 nm was achieved at 40 ℃ and a deposition rate of 1 Å/s, while an ultra-thin continuous Al film with a thickness of only 2 nm was obtained at a temperature below 0 ℃ and a reduced rate of 0.37 Å/s. The excellent interface sharpness and surface flatness of these epitaxially grown Al–Ge heterostructures hold promise for exploring topological quantum devices on Ge-based system. Moreover, the atomically flat surfaces and interfaces of the Al films enable the subsequent epitaxial growth of topological materials on top, as well as the fabrication of high-quality superconducting resonators.

## Supporting Information

1. Growth method
2. Methodology for quasi-1D FFT at the Al/Ge interface
3. Determination of Al/Ge interface width
4. Optimization of the growth rate in Recipe B

## Author Information

Jian-Huan Wang and Ding-Ming Huang contributed equally to this work.

*Corresponding Author: Jian-Huan Wang (wangjianhuan@baqis.ac.cn); H. Q. Xu (hqxu@pku.edu.cn) ; Jian-Jun Zhang (jjzhang@iphy.ac.cn).

**Notes**

The authors declare no competing financial interest.

**Acknowledgements**

This work was supported by the Natural Science Foundation of China (NSFC) (Nos. 62225407, 92565304, 12304101, 12304207), National Key Research and Development Program of China (2025YFE0217400), and the Innovation Program for Quantum Science and Technology (No. 2021ZD0302300).

# SUPPORTING INFORMATION

## Single-Crystalline Al/Ge Heterostructure with an Atomically Sharp Commensurate Interface

Jian-Huan Wang[1,2,*], Ding-Ming Huang[1,2], Han Gao[1], Yuan Yao[2], H. Q. Xu[1,3,*] and Jian-Jun Zhang[2,*]

[1]*Beijing Academy of Quantum Information Sciences, Beijing 100193, China*
[2]*Beijing National Laboratory for Condensed Matter Physics and Institute of Physics, Chinese Academy of Sciences, Beijing 100190, China*
[3]*Beijing Key Laboratory of Quantum Devices, Peking University, Beijing 100871, China*

Correspondence to: wangjianhuan@baqis.ac.cn (J.-H. Wang);
hqxu@pku.edu.cn (H. Q. Xu);
jjzhang@iphy.ac.cn (J.-J. Zhang.)

## Growth method

The Al/Ge heterostructures were grown on commercial epi-ready Ge(111) wafers using a dual-chamber solid-source MBE system with a base pressure of $1\times10^{10}$ mbar. The samples were first baked at 180 °C for 5 hours in a load-lock chamber to remove the moisture on the surface, and then transferred to the SiGe MBE chamber for the growth of Ge. Before growth, samples underwent degassing at 300 °C for 20 mins followed by deoxidation at 550 °C for 10 mins. Next, a 60-nm thick Ge layer was deposited on the samples utilizing a conventional two-step growth process to achieve an ideal surface for subsequent Al film growth. First, a 30-nm low-temperature Ge layer was deposited at 280 °C to cover the possible contamination and to recover the surface. Then, another 30-nm high-temperature layer was grown at 320 °C to obtain an atomic flat surface. During this two-step process, Ge was deposited at a fixed rate of 0.5 Å/s via electron beam evaporation.

After the growth of Ge layer, samples were cooling down and then in-situ transferred to the MBE chamber B for Al deposition. High-purity Al was evaporated onto the sample surface from a PBN crucible in a standard cold-lip Knudsen-type effusion cell. We conducted a systematic study on Al film growth by varying growth temperature and deposition rate, resulting in two optimized recipes. Recipe A involves depositing Al films of varying thicknesses at a fixed substrate temperature of 40 °C and at a growth rate of 1 Å/s, under which conditions the grown Al films exhibit optimal surface flatness. In contrast, Recipe B employs a substantially lower growth temperature (below 0 °C) and a reduced rate of 0.37 Å/s to deposit Al films. To minimize the sample temperature in Recipe B as much as possible, the backside of the as-Ge-epilayer-grown sample was thermally anchored to a cold plate for 3 hours, where the cold plate itself is connected to the MBE $LN_2$ cooling shroud via copper tape. Concurrently, in Chamber B, sample stage heater, beam flux monitor (BFM), and heaters of all effusion cell except the Al cell were turned off 72 hours in advance. After the sample was sufficiently pre-cooled, it was transferred into chamber B, and Al deposition commenced immediately. Throughout the entire growth process, the Type-K thermocouple reading on the chamber B substrate stage remained below its minimum measurable temperature of 0°C.

## Methodology for quasi-1D FFT at the Al/Ge interface

(a)

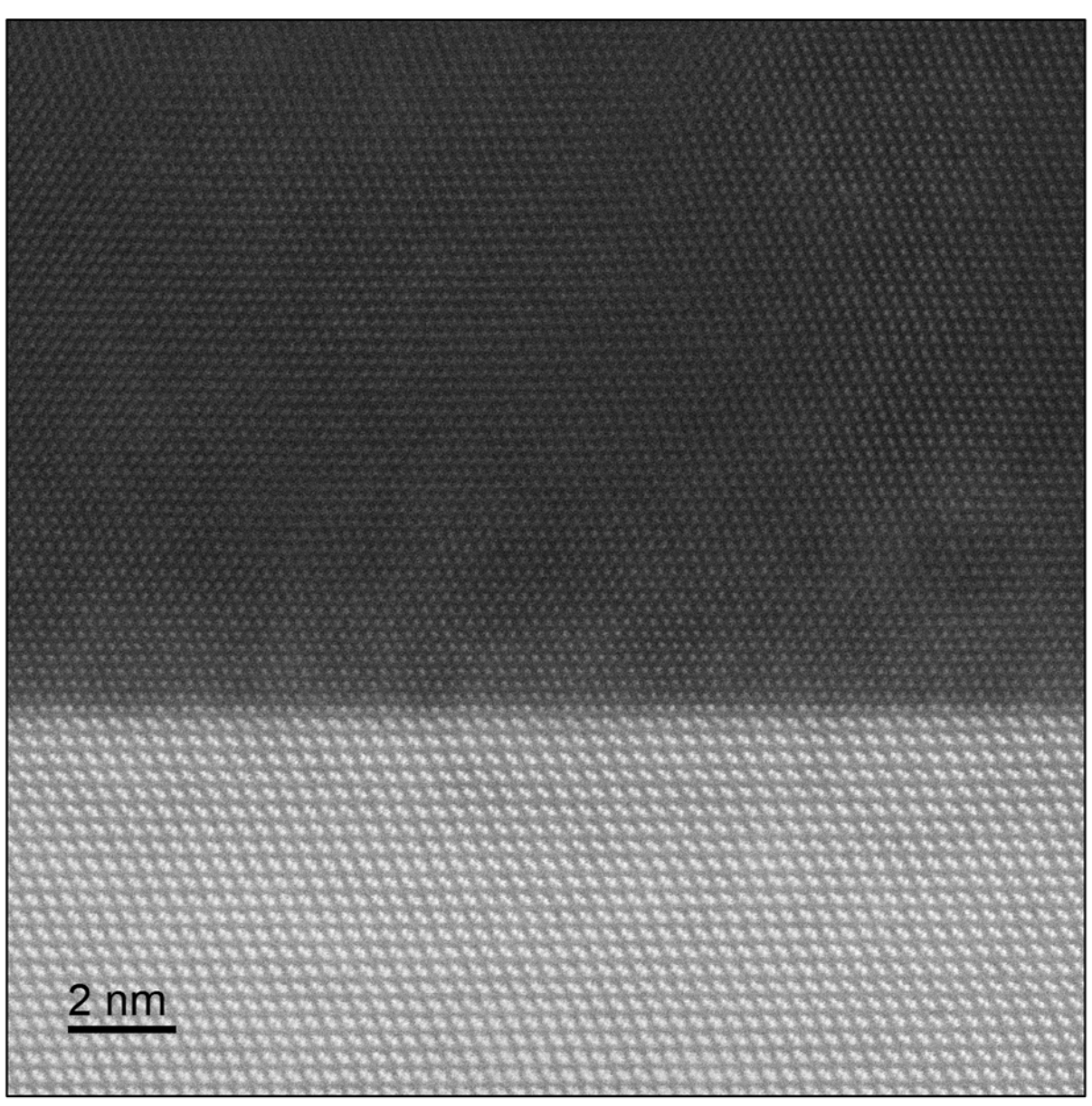

(b)

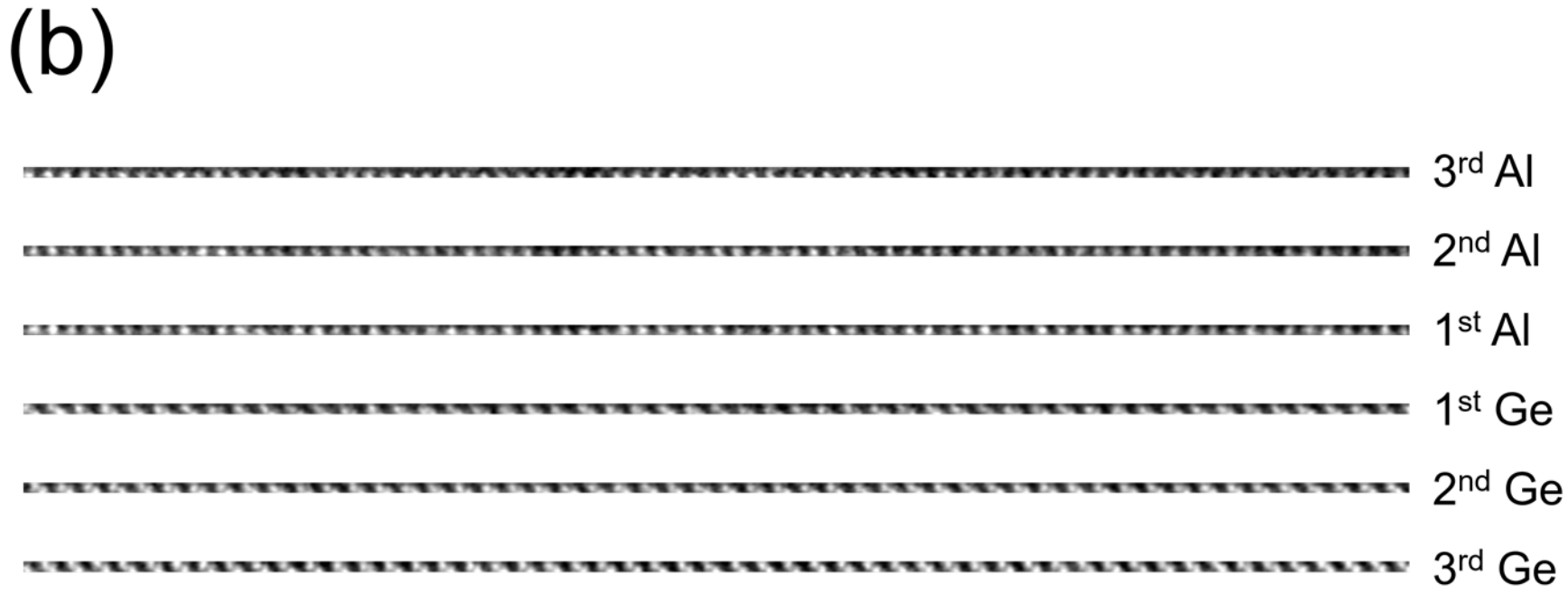

**Figure S1. Methodology for quasi-1D FFT at the Al/Ge interface.** (a) Low-magnification HAADF-STEM image of the Al/Ge heterostructure. (b) Atomic columns used for FFT, confined to narrow regions to exclude artifacts from adjacent layers.

To better resolve the periodicity of the atomic arrangement at the interface, a larger-scale HAADF-STEM image was selected for FFT analysis, as shown in Fig. S1(a). The specific atomic columns analyzed are shown in Fig. S1(b), where a narrow region was defined to ensure the exclusion of potential artifacts from adjacent atomic layers.

## Determination of Al/Ge interface width

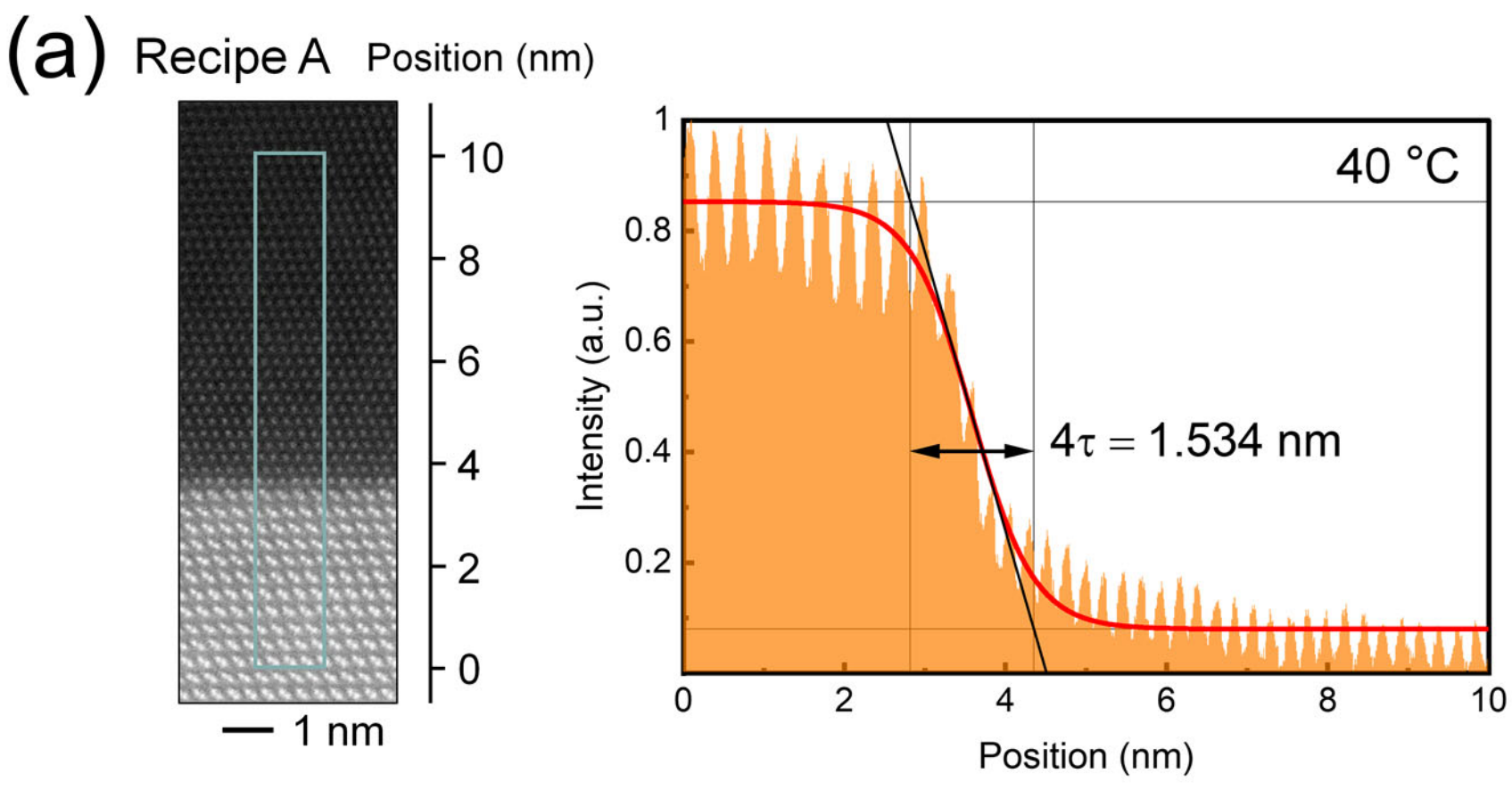

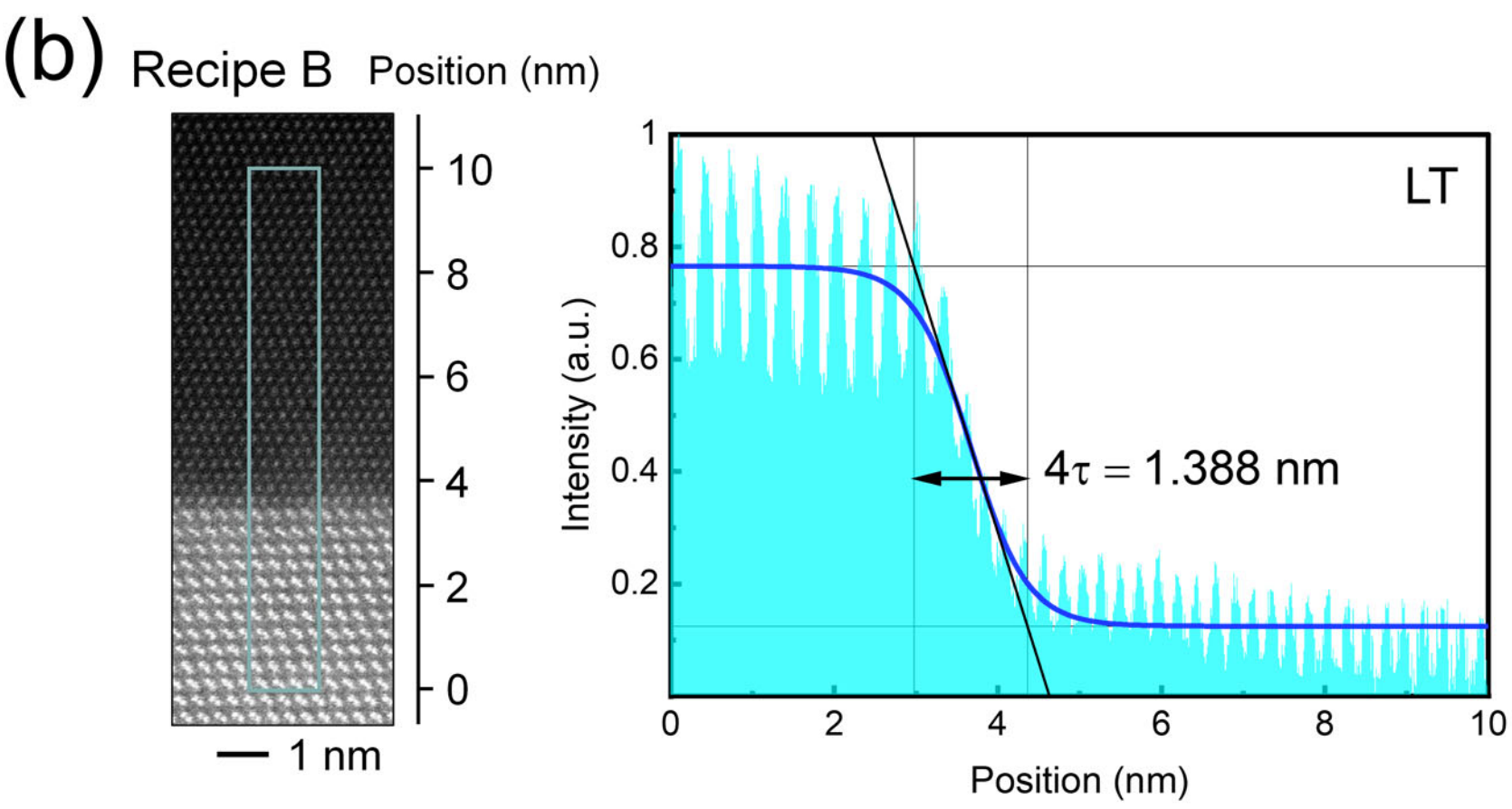

**Figure S2. Determination of Al/Ge interface width.** (a, b) Cross-sectional HAADF-STEM images (left) and their corresponding intensity profiles (right) acquired from the cyan frames in the films grown by (a) Recipe A and (b) Recipe B. Note that the cross-sectional HAADF-STEM image shown in (b) is the same as in Fig. 5(c). The solid curves in the profiles are sigmoid fits, with the interface width defined by the fitted parameter $4\tau$.

The interfacial widths of both Recipe A and Recipe B were quantified, as summarized in Fig. S2. The measured widths are 1.534 nm and 1.388 nm for the heterostructure grown by Recipe A and Recipe B, respectively. This comparison confirms that both interfaces are atomically sharp, with the interface width obtained by Recipe B being slightly narrower. We attribute this improvement to the suppressed surface atom migration and diffusion enabled by lower growth temperature in Recipe B.

## Optimization of the growth rate in Recipe B

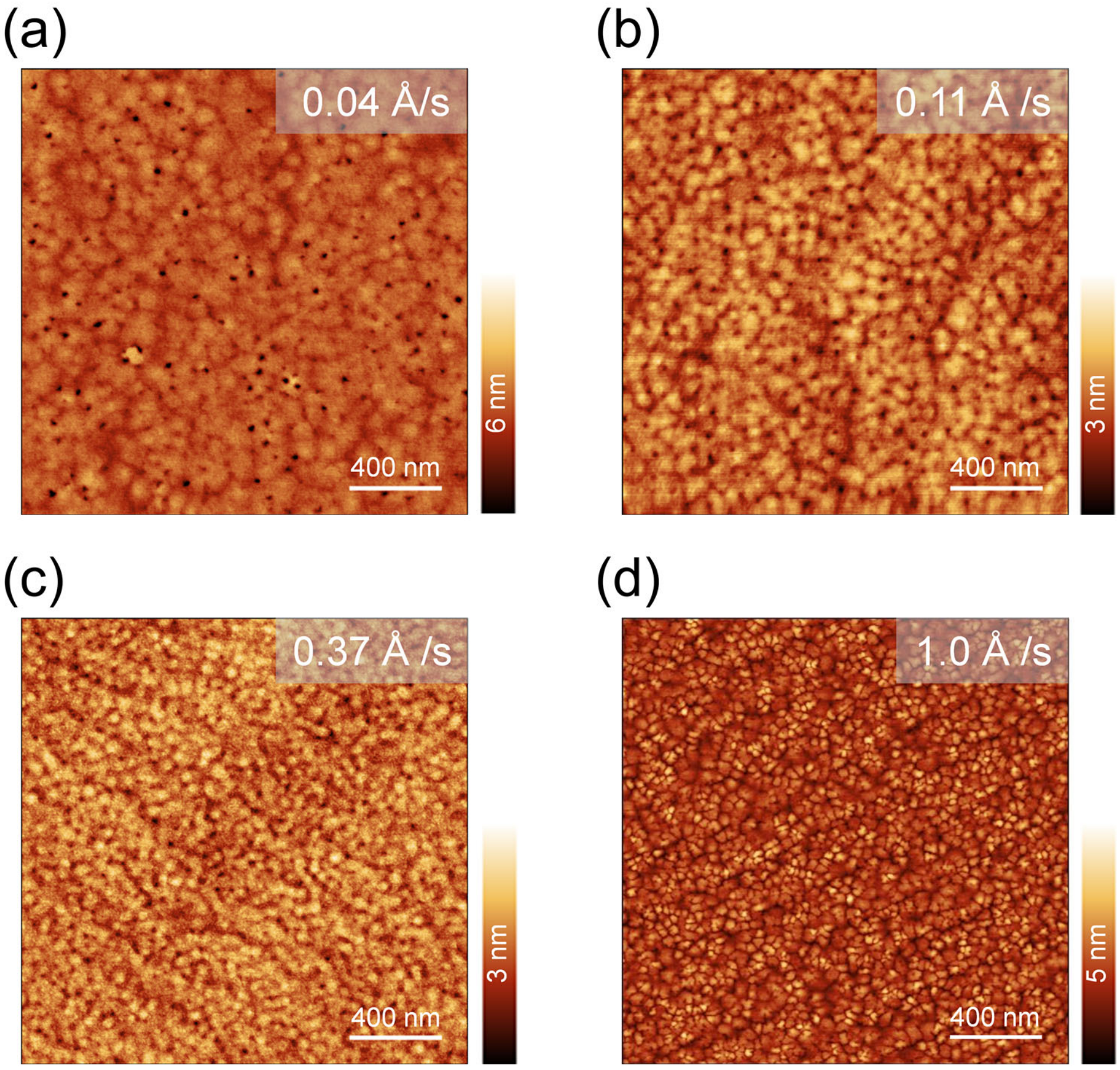

**Figure S3. Optimization of the growth rate in Recipe B.** (a-d) AFM images (2 μm × 2 μm) showing 20 nm-thick Al films grown at rates of (a) 0.04 Å/s, (b) 0.11 Å/s, (c) 0.37 Å/s, (d) 1.0 Å/s.

Figure S3 (a-d) presents the surface morphologies of 20-nm-thick Al films grown at low-temperature conditions with Al growth rates of 0.04 Å/s, 0.11 Å/s, 0.37 Å/s, and 1.0 Å/s, respectively. Compared to the results obtained by Recipe A (as shown in Fig. 1(b) and Fig. 4(d)), the Al films grown at these lower-temperature conditions exhibit a characteristic island-like surface. For Recipe B, the growth rate was selected based on minimal surface roughness. The film grown at 0.37 Å/s (Fig. S3(c)) was confirmed to possess the optimal flatness with an RMS roughness of 0.326 nm.